\documentclass[fleqn,usenatbib]{mnras}

\usepackage{newtxtext,newtxmath}

\usepackage[T1]{fontenc}

\DeclareRobustCommand{\VAN}[3]{#2}
\let\VANthebibliography\thebibliography
\def\thebibliography{\DeclareRobustCommand{\VAN}[3]{##3}\VANthebibliography}

\usepackage{graphicx}	
\usepackage{amsmath}	
\usepackage{amssymb}	
\usepackage{subcaption} 
\usepackage[normalem]{ulem}

\newcommand{\taus}{\tau_{\mathrm{s}}}

\newcommand{\vu}{\mathbf{u}}

\title[Dynamics of Dusty Vortices II]{Dynamics of Dusty Vortices II: Stability of 2D dust laden vortices}

\author[F. Lovascio et al.]{
	Francesco Lovascio,$^{1,2}$\thanks{E-mail: francesco.lovascio@ens-lyon.fr}
	Sijme-Jan Paardekooper,$^{2}$\thanks{E-mail: s.j.paardekooper@qmul.ac.uk}
	Colin McNally
	\\
	$^{1}$Univ Lyon, Univ Lyon1, Ens de Lyon, CNRS, Centre de Recherche Astrophysique de Lyon UMR5574, F-69230, Saint-Genis,-Laval, France.\\
	$^{2}$School of Physics and Astronomy, Queen Mary, University of London, , E1 4NS, UK
}

\date{Accepted XXX. Received YYY; in original form ZZZ}

\pubyear{2015}

\begin{document}
\label{firstpage}
\pagerange{\pageref{firstpage}--\pageref{lastpage}}
\maketitle

\begin{abstract}
Vortices have long been speculated to play a role in planet formation, via the collection of dust in the pressure maxima that arise at the cores of vortices in protoplanetary discs. The question remains however: as dust collects in the core of a vortex, when does that vortex remain stable and able to collect further dust, and when and why does it break up? We study this question by running high resolution 2D simulations of dust laden vortices. By using the terminal velocity approximation in a local shearing box it was possible to efficiently run simulations of back-reacting dust in a gas at high resolution. Our results show how the stability of 2D dust laden vortices in protoplanetary discs depends on their size relative to the disc scale height, as well as the dust coupling. We find small vortices with semiminor axis much smaller than the scale height to be stable for the duration of the simulations ($t>2000$ orbits). Larger vortices, with semiminor axis smaller than but of order of the scale height, exhibit a drag instability after undergoing a long period of contraction where the core becomes progressively more dust rich. The lifetime of these vortices depends on the dust size, with larger dust grains causing the instability to occur sooner. For the size ranges tested in this paper, $\mu$m to mm sized grains, vortices survived for several hundreds of orbits. The result implies that the stability of vortices formed by vertical shear instability and zombie vortex instability, or the breakup of larger vortices through hydrodynamic instabilities, is affected by the presence of dust in the disc. The lifetimes observed in this paper, while shortened by the presence of dust for larger vortices, were still long enough to lead to considerable dust enrichment in the vortex cores.
\end{abstract}




\section{Introduction}

Many barriers exist for the growth of dust into planets, yet we observe a zoo of exoplanets, with a large variety possibly still concealed by observational limitations. Reconciling observations with our theoretical understanding has been a standing challenge in theoretical planet formation, especially with the advent of extensive exoplanet surveys. It is known that traditional growth models for kilometre-sized planetesimals, that rely on collisions between dust particles, struggle to overcome growth barriers like the bouncing, fragmentation, and drift barriers. At the bouncing barrier, collisions between particles are more likely to lead to bouncing than sticking \citep{Zsom_2010}. This occurs at particle sizes $a$ of order $a\sim1\mathrm{mm}$. If particles pass the bouncing barrier, when sizes ranging from $a\sim1~\mathrm{cm}$ to $a\sim1~\mathrm{m}$ they reach the fragmentation barrier. Once particles reach the fragmentation barrier, impacts become more likely to break the particles into smaller pieces, halting growth \citep{Blum_2008}. This occurs because the energy of the collisions scales with the mass of the object, while the strength can only scale with the cross-sectional area, meaning that, as grains grow collisions become more and more likely to have more energy than the structural strength of the colliding grains can withstand, leading to the collision being a fragmenting collision. At sizes $a\sim1\mathrm{m}$ particles also encounter the drift barrier. At the drift barrier, particles become sufficiently affected by the gas drag that the drift timescale becomes comparable to the growth timescale, at which point grains drift into the central star before they can grow by a significant amount \citep{Weidenschilling_1977}. Finally, it is worth pointing out that such oligarchic growth models have been shown to be unable to reproduce angular momentum distributions of solar system bodies \citep{Johansen_2015_book}.

For these growth barriers to be overcome a mechanism to drive rapid growth is required.  Some proposed alternative planet formation models require instabilities in the protoplanetary disc in order to rapidly grow planets. For gas giant planets, it is known that Gravitational instabilities (GI) are able to collapse the outer regions of a disc into large clumps if the Toomre Q \citep{Safronov_1960,Toomre_1964} is of order unity and the disc can cool efficiently enough \citep{Rafikov_2005,Kimura_2012}. Gravitational instabilities in massive discs have been shown to reliably produce massive gaseous planets from gravitationally unstable discs. Planets formed by GI have a metallicity similar to that of the disc, which does not match solar system observations, suggesting that solar system gas giants did not in fact form by GI \citep{Johnson_2013}. While self gravitating discs have been observed, GI is not sufficient for forming the variety of gas giants observed, implying that at least some must form by runaway gas accretion onto a rocky core, still requiring a fast formation pathway for large rocky planetesimals.

GI for the dust component of the disc have been invoked for fast planetesimal formation \citep{1969edo..book.....S, 1973ApJ...183.1051G}, but was thought to require an extremely laminar disc \citep{1993prpl.conf.1031W}. It was later shown that in fact turbulence such as generated by the Magneto Rotational Instability \citep{osti_4232891,1960PNAS...46..253C,1991ApJ...376..214B} could actively help to concentrate in particular large boulders to an extent that GI could set in \citep{2006ApJ...636.1121J}. Additionally, (magneto-) hydrodynamic instabilities can act to create dust traps which accelerate planet formation \citep{Lyra_2012, Riols_2020}. Therefore, with some external help to concentrate solids, GI could be a plausible fast growth mechanism for solid bodies. 

The streaming instability \citep{Youdin_2005} is one of the leading theories for planetesimal formation by concentrating solids enough for GI to take over. This instability is a member of a wider family of resonant drag instabilities as described by \citet{Squire_2018}. Drag instabilities occur whenever there is a dust species drifting through a gas species. Drag resonances cause dust over-densities to act as traffic jams causing an exponential growth in local dust densities. In the streaming instability the dust streaming is due to the radial drift of the dust particles. This radial drift is caused by the headwind felt by the dust particles, as gas orbits the star at a sub Keplerian velocity due to being partly pressure supported \citep{Weidenschilling_1977}. While the streaming instability constitutes one of the most promising mechanisms for overcoming the barriers to planet formation, some issues remain. For example, the streaming instability seems to become less efficient when considering multiple dust species \citep{Krapp_2019} and be constrained to very high wave number regimes when considering a continuum in dust size distribution \citep{PSI1, PSI2, PSI3}. Turbulent diffusion can also can act to damp, at least the linear phase of the streaming instability \citep{Lin_2021,Umrhan_2020}, especially in the case of dust size distributions \citep{PSI3}.  
Other processes can also cause dust to stream through the gas in the disc. Notably, the vertical settling instability has a drag instability take place as dust settles towards the disc midplane \citep{2018MNRAS.477.5011S}. These drag instabilities however, appear to be insufficiently efficient at collecting dust for them to be a strong candidate as a planet formation mechanism \citep{Krapp_2020}. The vertical settling instability has not been studied as extensively as the streaming instability, but also does not show great promise for planet formation \citep{Krapp_2020}, as it does not seem to be able to form large dust clumps.  

Vortices have been speculated to play a role in planet formation, as an alternative mechanism to form local dust over densities. In protoplanetary discs, vortices have a pressure maximum at their core. Dust collects at pressure maxima in scenarios where dust is streaming through a fluid, the pressure maximum at a vortex core is therefore able to collect dust from its surroundings \citep{Barge_1995}. Dust in a vortex continuously drifts towards the vortex centre, as with other examples of dust drift, the drift velocity of the dust depends on the strength of the dust-gas coupling, referred from here on as the Stokes number $\mathrm{St}$. Particles with $\mathrm{St} \ll 1$ are well-coupled to the gas, while particles with $\mathrm{St}\sim 1$ have the fastest drift velocity. In a protoplanetary disc (PPD), the Stokes number $\mathrm{St}=\taus\Omega$ is set by the product of stopping time $\taus$ to orbital frequency $\Omega$. The stopping time introduced above is the time taken for a grain to reach terminal velocity in a fluid. Recent work by \cite{Gerbig_2020} shows how what is needed for planetesimal formation to be triggered is just a sufficient dust overdensity for gravitational collapse to overcome turbulent diffusion and tidal shear. The rate at which dust collects in vortex cores depends on the Stokes number of the dust particles, which opens up the possibility for vortices to act as a mechanism to size sort dust. Dust size sorting is important in the context of planet formation, as there is evidence to suggest that the streaming instability is sensitive to the dust size distribution function \citep{PSI3}.

Vortices have plenty of opportunities to form in PPDs. Simulations of PPDs, both local and global, have shown vortices appearing throughout discs, brought about by various hydrodynamic and magneto-hydrodynamic (MHD) instabilities \citep{Lovelace_2014,Marcus_2015}. Steep density gradients, for example caused by a massive planet opening up a gap in the disc, are known to be prone to the Rossby Wave Instability \citep{Lovelace_1999}, giving rise to large scale vortices as seen in \cite{Koller_2003} and in 3D in \cite{Lin_2012}. These vortices are hydrodynamically stable, meaning that, in isolation, their lifetime is only determined by the viscous damping \citep[see e.g.][]{Fu_2014}. Other effects can also lead to the breakup of vortices, including cooling \citep{Fung_2021, Les_2015, Lobo-Gomes_2015, Rometsch_2021} and shocks in spiral planet induced wakes \citep{Hammer_2021}. These very large vortices typically have lifetimes of the order of thousands of orbits for turbulent viscosities $\alpha<10^{-4}$ in optimal conditions, though the lifetimes can be significantly shortened by the aforementioned effects. Vortices can also form due to the vertical shear instability (VSI) \citep{Richard_2016}, as temperature gradients in the disc drive vertical mixing. While early results suggested that the vortices formed this way would be short-lived, more recent results have found the VSI capable of producing larger vortices of size similar to the scale height with lifetimes of order hundreds of orbits \citep{Manger_2018}. Finally, vortices may propagate through a disc via the zombie vortex instability, which allows for vortices to form in the disc if a sufficiently large vorticity perturbation is present in the disc \citep{Marcus_2015}. Due to the subcritical baroclinic instability, even small anticyclonic vortices can be amplified into large anti-cyclones and be sustained for long periods \citep{Lesur_2010}.

For vortices to aid in the formation of planets, vortices in protoplanetary discs need to be sufficiently long lived to collect and hold dust in their cores, or at least size sort and increase dust concentration sufficiently to kick off the SI. When not considering dust, vortices in protoplanetary discs have been shown to have long lifetimes \citep{Lesur_2010}, but the hydrodynamics of dusty core vortices are not fully understood. Several studies have shown instabilities in vortices as the vortices become more dust enriched \citep{Fu_2014,Crnkovic-Rubsamen_2015,Surville_2019}. A few different mechanisms have been suggested for this instability, from drag  instabilities to a heavy core instability \citep{Chang_2010}. 
All studies above were carried out in two spatial dimensions. While the full 3D problem needs attention, it is worth fully understanding the 2D stability problem first.
Theoretical work has been done on the stability of dusty vortices in 2D, where \citet{Railton_2014} showed that dusty vortices of aspect ratio less than 4 would go unstable, while \citet{Chang_2010} derived a linear instability affecting all shear free dusty core vortices analogous to the Rayleigh-Taylor instability. In this 'heavy core' instability the centrifugal force in the vortex causes there to be a force acting against the density gradient in the fluid, a configuration which is Rayleigh-Taylor unstable \citep{Sipp_2005}. \citet{Railton_2014} showed however that the heavy core instability can only arise in a shear free vortex, which for a Keplerian shearing background can only occur for a vortex aspect ratio of 7. Both \citet{Railton_2014} and \citet{Chang_2010} treat the case of a perfectly coupled dust fluid, but to study drag instabilities Stokes numbers greater than 0 need to also be considered. This leads to a tension in the literature; dusty vortices have been observed to go unstable in computational studies, with two proposed causes for the instability the heavy core instability and drag instabilities. In this paper we run nonlinear simulations of dusty vortices with perfect and imperfect coupling to study their stabilities in an attempt to resolve this conflict in the literature. 
In this paper, the numerical setup section (section \ref{sec:nummethods}) covers the setup used in the simulations in this paper, describing the initial conditions, equations solved, code used, and numerical experiments run. The results section (section \ref{sec:results}) covers the results obtained in the numerical experiments, attempting to break up the results into broader families of cases: small vortices and large vortices. The discussion section (section \ref{sec:discussion}) discusses the origins of the observed vortex instability and its implications for vortices in protoplanetary discs, as well as how this work ties in with other work done by the authors and the wider community. Finally in the conclusion section (section \ref{sec:conclusion}) the results of the paper are summarised.

\section{Numerical Setup}
\label{sec:nummethods} Studying dusty vortices poses a significant computational challenge. Evolution equations for a dusty gas are well known to be expensive to solve. This is because the dust-gas coupling terms are stiff \citep{Lovascio_2019}, meaning that they are subject to spurious mode growth when being solved numerically \citep{iserles_book}. This phenomenon exhibits itself as qualitatively incorrect solutions to the problem when numerically solving with too coarse time resolution. Additionally dusty vortices can have small scale substructure, requiring high spatial resolution for consistent behaviour \citep{Surville_2019}, further exacerbating the challenges posed by the stiffness of the governing equations. We employ several approaches to mitigate these difficulties. All our simulations are run in the inviscid terminal velocity approximation first described by \citet{Laibe_2014} which simplifies the problem by considering the dust to be at terminal velocity with respect to the gas, reducing the number of equations to solve. The approximation we use however is that described in \cite{Lin_2017}, which further simplifies the approximation from \citet{Laibe_2014} by taking a locally isothermal equation of state for the gas, allowing for the feedback from the dust on the gas to be expressed as a cooling term in the 'energy equation' of the dust-gas mixture. While the cooling term from \cite{Lin_2017} can be numerically stiff, its stiffness decreases with increasing dust coupling \citep{Lovascio_2019}, meaning that the small particle sizes most relevant to the study of young protoplanetary discs \citep{Trotta_2013,Hull_2018} and the ones best modelled by the terminal velocity approximation \citep{Laibe_2014, Lin_2017, Lovascio_2019} do not cause the cooling term to be excessively challenging to integrate. The scheme we employ, \textsc{fargo-fls} is a modification on the \textsc{fargo3d} hydrodynamics code, which uses the unconditionally stable explicit integration scheme from \citet{Meyer_2012} and \citet{Meyer_2014} to advance the dust cooling term. The scheme is tested and showcased in our previous paper, \cite{Lovascio_2019}. The choice of a modified vertically unstratified shearing box geometry allows us to reduce the size of the computational domain and cheaply gain more resolution for the vortex and neglect the vertical direction as unstratified discs do not depend on the vertical direction. We apply the shearing box approximation to the locally isothermal terminal velocity approximation as described in \citet{Lin_2017},
\begin{gather}
    \partial_t\rho +\nabla\cdot\left(\rho\vu\right)=0 \\
    \partial_t\left(\rho\vu\right)+\nabla\cdot\left(\rho\vu\vu\right)=\rho\Omega^2\left(2q\mathbf{y}-\mathbf{z}\right) - 2\Omega\hat{\mathbf{z}}\times\rho\vu\\
    \partial_t P + \nabla\cdot\left(P\vu\right)=\nabla\cdot\left[\taus \left(1-\frac{P}{c_s^2\rho}\right)\nabla P\right] + \Omega^2 \rho \vu\cdot\left(2q\mathbf{y}-\mathbf{z}\right)
\end{gather}
It is important to note that in this work the coordinate system used is the same as the \textsc{fargo3d} code. In this coordinate system the $x$ direction is the azimuthal direction and $y$ the radial. In order to keep the problem 2D, the initial conditions are taken to be independent of the vertical coordinate $z$, so that all vertical gradients vanish. This then remains true for the whole simulation, and we can thus use only a single computational cell in $z$.

The simulations differ from a conventional shearing box, where all the boundaries are either periodic or shear-periodic, in that a wave killing boundary layer is implemented in the $\mathbf{\hat{y}}$ direction (the radial direction when considering the whole disc). Vortices in discs emit density waves \citep[e.g.][]{2010ApJ...725..146P}, which propagate through the shearing box and appear on the other side. Wave-killing boundaries help to reduce the pollution of the computational domain by these waves. The damping boundary we implement uses a smooth $\sin^2$ mask defined as follows,
\begin{equation}
\begin{split}
    Q_{boundary}= Q_{simulated}+\\
    \sin^2\left(\frac{\pi y}{L}\right)\beta\tau\cdot\left(Q_{background}-Q_{simulated} \right)
\end{split}
\end{equation}
where $Q_{boundary}$ is the value given in the boundary, $Q_{simulated}$ and $Q_{background}$ are the value calculated by the solver and the background shearing box value respectively, $\frac{y}{L}$ is the distance from the edge of the damping region over the size of the damping region, $\tau$ is the timestep and, $\beta$ is a parameter to tune the strength of the damping. 
To initialise a vortex in our simulations we set the velocity field in the domain to a Kida \citep{Kida_1981} vortex. A Kida vortex is a steady-state solution of a shearing flow with an elliptical vorticity patch,
\begin{gather}
    u_x =\begin{cases}
    \frac{3}{2}\Omega y &\mbox{outside the vortex}\\ 
    -\frac{q^2}{1 + q^2}\omega y &\mbox{inside the vortex}
    \end{cases}\\
    u_y =\begin{cases}
    0 &\mbox{outside the vortex}\\ 
    -\frac{1}{1 + q^2}\omega x &\mbox{inside the vortex.}
    \end{cases}
\end{gather}
The vorticity $\omega$ is then related to the aspect ratio of the vortex $q=\frac{a}{b}$ (where $a$ is the semimajor axis of the vortex and $b$ the semiminor) and the Keplerian shear $\omega_\mathrm{K}$ by 
\begin{equation}
    \frac{\omega_\mathrm{K}}{\omega}=\frac{q^2-q}{1+q^2}.
\end{equation}
As there is no steady state dusty vortex solution and the slowly varying solution requires a boundary value problem to be solved to initiate the pressure field \citep{Railton_2014}, we do not attempt to initialise the pressure to a Kida-like pressure profile, we rather set a flat pressure profile, with a corresponding flat density profile such that the dust to mass ratio is $f_{d(vortex)}$ inside the vortex and $f_{d(background)}$ elsewhere, with a sharp transition at the edge of the vortex. Initialising the vortices this way means that there will be an initial relaxation period where the vortex contracts towards a quasi-stable configuration, that will slowly change due to further dust loading. For all experiments run in this paper, the vortices survive this adjustment period without major disruption. This is illustrated in figure \ref{fig:DV_Life-of-a-small-Vortex}, far left panel, where the vortex has not yet reached the slowly evolving configuration after $\Omega t =500$, as indicated by the wavy pattern in dust-to-mass ratio. After $\Omega t=2000$, a smooth, slowly varying dust to mass configuration has been reached (middle panel of figure \ref{fig:DV_Life-of-a-small-Vortex}). 
In this paper we study the different behaviour of large and small vortices loaded with dust of different coupling. In the shearing box setup we use, to vary the vortex size it is not required to vary the extent of the vortex in simulation units, due to the scaling properties of a shearing box varying the gas sound speed, $c_s$ has the same effect as changing the size of the box in terms of the pressure scale height $H$. We assume the gas to be isothermal, this makes the scale height of the disc $H$ 
\begin{equation}
	H \propto c_s.
\end{equation}
Keeping all other things equal, a run with the sound speed reduced by a factor of 2 corresponds to a vortex that is twice as large in terms of $H$ compared to the original simulation. As the behaviour of vortices in PPDs is set by their radial size relative to the scale height \citep{Shen_2006}, we explore a parameter space from small vortices, where $b \ll H$, to large vortices, where $b \sim H$.  Exploiting this symmetry of the problem it is possible to eliminate resolution effects potentially damping instabilities. In our simulations the semi-minor axis of both large and small vortices is resolved by the same number of computational cells, this is because the size of the vortices is adjusted by scaling the gas sound-speed and thus the relation between $b$ and $H$. To study the relationship between between vortex stability and dust grain size we make use of \textsc{fargo3d}'s restart function, allowing us to restart a simulation with different parameters.

\subsection{Initial conditions}
In the numerical experiments run in this paper, the variables explored are the vortex size and dust coupling, the other initial conditions were kept fixed throughout all the simulations. The vortices were initialised with $a=1.0$, and $b=0.2$ (in code units) making their aspect ratio $q=5$. This aspect ratio was chosen as it is known to produce hydrodynamically stable vortices \citep{Railton_2014}. As the dust loading of the vortices happens over a long timescale, the vortex is pre-loaded with dust in the initial conditions, such that the dust to gas ratio is 
\begin{gather}
    f_\mathrm{d} =\begin{cases}
    0.021 &\mbox{outside the vortex}\\ 
    0.4 &\mbox{inside the vortex}
    \end{cases}.
\end{gather}
The sound speed, $c_s$ (units of $L/\Omega$ where $L$ is the box length) and stopping time $\taus$ are then varied in the simulations to study the effect of vortex size and dust coupling on vortex stability.

\subsection{Resolution and convergence testing}
All the 'resolved' simulations run in this paper are run at a resolution of $1440\times1440$ grid cells, which grants a resolution of about 100 cells across the semi-minor axis of the vortices. Simulations of resolution up to $2096\times2096$ were run, showing that the behaviour of the vortices remain qualitatively unchanged past resolutions of about $1200\times1200$. An important limitation in terms of resolution appear with regards to the box size used. If the box is too small, then  the vortex may interact with its self through the periodic boundary in the azimuthal direction, or with the wave damping boundaries in the radial direction. When this happens, the dynamics of the vortex-vortex interaction become important and new instabilities can arise. This kind of setup with vortex "chains" is not necessarily uninteresting or unphysical, however new instabilities are possible in this scenario (further discussed in section \ref{sec:interactions}) making everything much messier. This problem is especially serious for the smaller vortices, due to the method chosen to re-scale the problem. The larger sound-speed used in the small vortex simulations to re-scale the vortex size reduces the physical size of the box at the same time, resulting (at any given box size) in a box of smaller physical dimensions. Choosing boundaries at a distance of $x>2c_s/\Omega$ was found to be sufficient to prevent vortex self-interaction.

\section{Results}
\label{sec:results}
Using the numerical method described in section \ref{sec:nummethods}, we test the relationship between the vortex size and dust coupling to study instabilities in dusty vortices as seen by \cite{Crnkovic-Rubsamen_2015}, \cite{Surville_2019}, and \citet{Fu_2014}. We run a variety of initial configurations, varying box size, vortex size and dust coupling. 

Our results can be summarized as follows. Instabilities are discovered in large dust loaded vortices as well as very tight vortex chains. To determine the nature of these instabilities we re-run the unstable phase of the evolution with perfectly coupled dust, this eliminates drag driven instabilities. In large vortices the instability vanishes when considering perfectly coupled dust, indicating a drag instability. In the case of vortex chains on the other hand, the instability evolves the same way regardless of dust coupling indicating a hydrodynamic instability. Below, we consider these aspects in more detail. 

\subsection{Small vortices: b << H}
Several factors known to affect the stability of vortices in gas flows similarly affect dusty vortices. The aspect ratio is probably the most important of these: no vortex of aspect ratio below 3 can be stable, this extends to dusty vortices \citep{Railton_2014}. In our simulations the aspect ratio was initially set to $q=5$, but as the vortex contracts due to dust moving towards the core, the aspect ratio does change. The vortices also stopped being exactly elliptical making the aspect ratio be less useful in describing the vortices. An additional parameter we find to be very important is the size of the vortex. The initial $b:H$ ratio in the small vortices ranged from $10^{-1}$ to $4\cdot10^{-2}$. This value, like $q$, changes as the vortices contract, becoming somewhat smaller, depending on $\taus$. We find small vortices to be stable. In our simulations, they survived for the duration of our simulations $\Omega t=4500$ at all dust stopping times tested, with no signs of instabilities appearing. This should be compared to the typical growth rates as found by \cite{Railton_2014} of $10^{2}\Omega$, and to the typical time scale for instability as seen in \cite{Crnkovic-Rubsamen_2015}, $\sim 100$ local orbits. Therefore, our small vortices are not affected by these instabilities. This is expected for \cite{Railton_2014}, since these instabilities are 3D in nature, but more surprising in view of previous numerical results obtained in 2D \cite{Fu_2014, Crnkovic-Rubsamen_2015,Surville_2019}.

\begin{figure*}
	\centering
	\includegraphics[width=\textwidth]{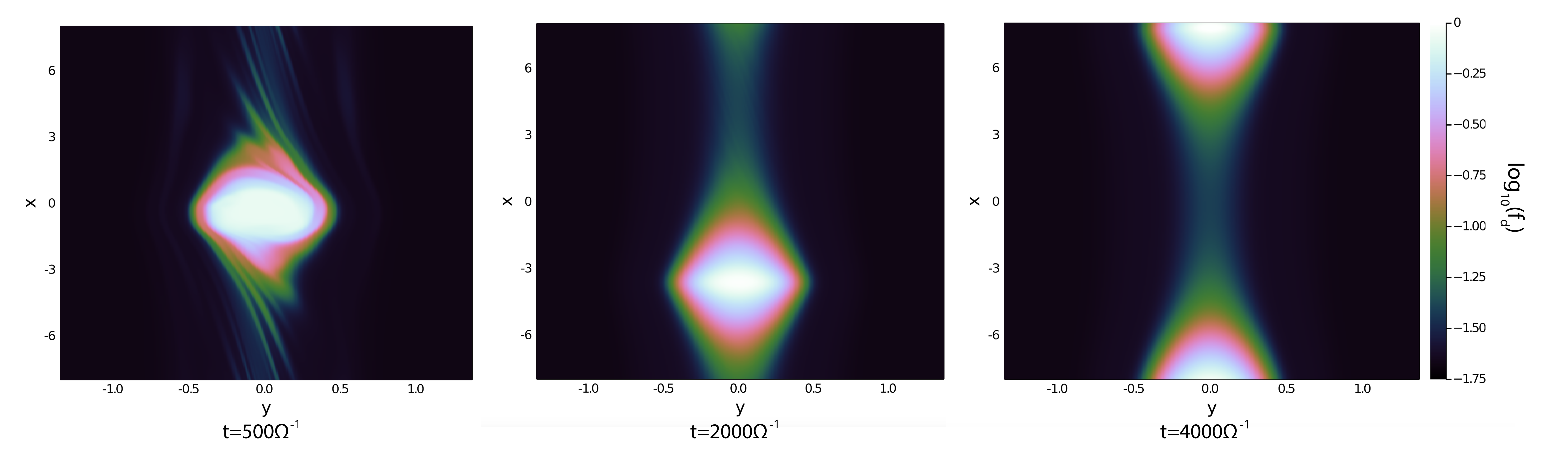}
	\caption{Figure showing the evolution of a small dusty vortex ($c_s=5L/\Omega$) with low dust coupling ($\Omega\tau=10^{-3}$). Plots show log dust fraction. The vortex is allowed to evolve under the effect of dust drag, showing no sign of instability at any point. The first panel on the left shows the end of the initial adjustment period when the vortex is ``looking for" a slowly evolving pressure profile.}
	\label{fig:DV_Life-of-a-small-Vortex}
\end{figure*}

The time evolution of small vortices can be seen in figure \ref{fig:DV_Life-of-a-small-Vortex} where a small vortex (using $c_s=5.0$) is allowed to evolve and no unstable configuration is reached. The left panel shows some residual swirly structures from the initial adjustment to a quasi-steady state. While we show the dust fraction, the extent of the vortex in terms of vorticity is well captured by the white region of largest dust fraction in the plots. Note that the dust fraction in the core of the vortex is almost unity, which makes this vortex extremely heavy with dust. Despite this, the vortex remains stable until $\Omega t=4500$. This suggests that small, dusty vortices are at least long lived.

\subsection{The life-cycle of a large dusty vortex}
\begin{figure*}
	\centering
	\includegraphics[width=\textwidth]{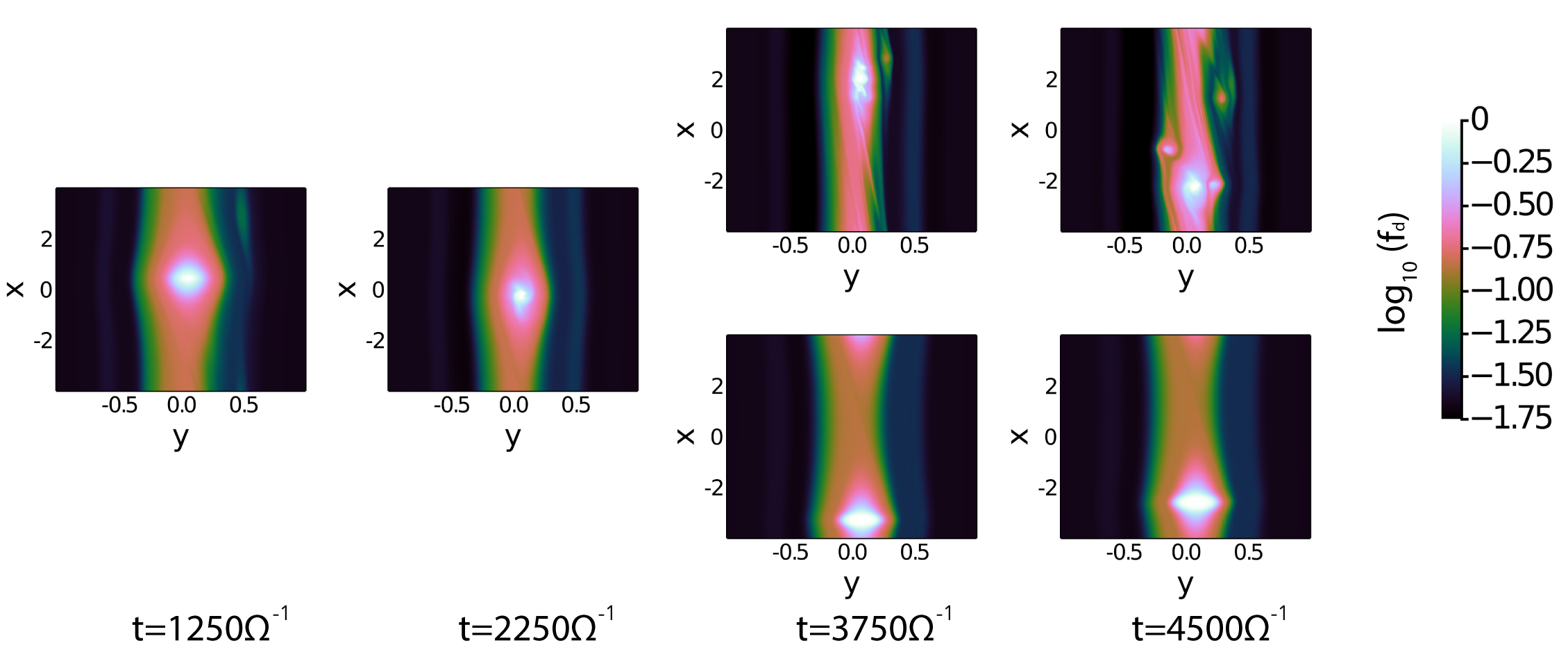}
	\caption{Figure showing the evolution of a dusty vortex. Plots show dust to mass ratio.The vortex is allowed to evolve under the effect of dust drag (left half) until the first sign of instability is shown ($\Omega t=2250$). At this point the simulation is checkpointed, allowing for another simulation to be launched where the dust feedback is disabled by setting $\taus=0$ and hence $C_d =0$ (bottom right row). The top row with dust feedback continues to exhibit instability until the vortex breaks up. The bottom row where the dust feedback has been disabled returns to a stable configuration, showing that the instability is a drag instability.}
	\label{fig:DV_Life-of-Vortex}
\end{figure*}

Two otherwise identical vortices can have vastly different lifetimes when differing only in physical size. The two vortices in Figures \ref{fig:DV_Life-of-a-small-Vortex} and \ref{fig:DV_Life-of-Vortex} are set up identically other than that one has $b\approx H$ ($b/H=0.2$), (Figure \ref{fig:DV_Life-of-Vortex}) and the other $b\ll H$ ($b/H=4\cdot10^{-2}$) (Figure \ref{fig:DV_Life-of-a-small-Vortex}). As the dust piles up into the core of the vortices, the smaller vortex rapidly finds a steadily contracting state, which remains stable till the end of the simulation at $\Omega t=4500$. For the larger vortex on the other hand the steadily contracting configuration is not stable, the vortex starts intermittently exhibiting spiral patterns (Figure \ref{fig:DV_Life-of-Vortex} panel 2), eventually breaking up into multiple smaller vortices (Figure \ref{fig:DV_Life-of-Vortex} upper panels 3 \& 4). If sufficient time is allowed to pass, the vortex dissipates until only a dust ring is left. In our simulation at the time of the instability appearing, the vortex aspect ratio was approximately 10, but the instability did not appear to be related to aspect ratio, but rather dust to gas ratio, as more dust loaded vortices went unstable earlier, when aspect ratios had not deviated significantly from the initial value of 5.

\begin{figure}
	\centering
	\includegraphics[width=0.9\columnwidth]{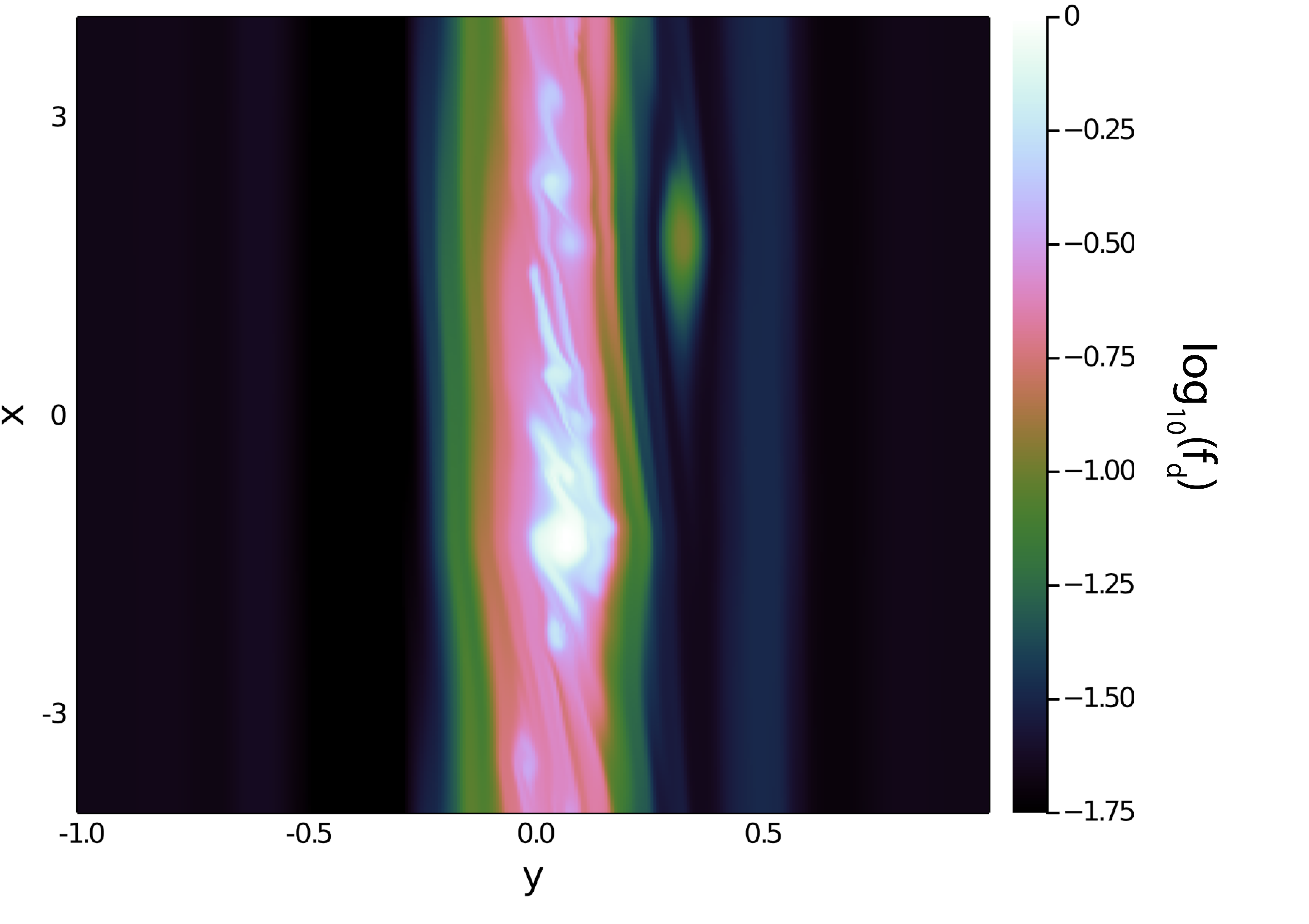}
	\caption{Plot showing dust ring and small vortices left at the end of the life of a large vortex ($c_s=1.0$, $\Omega\taus=10^{-3}$) at $\Omega t=5000$.}
	\label{fig:DV-ring}
\end{figure}

For large dust laden vortices with high dust concentrations in the core ($f_d \approx 1.0 $) this evolution appears inevitable, while different initial conditions can act to change the time taken for the instability to first appear, the steadily contracting configuration seems to be unstable at high dust to gas ratios. Several simulations were initiated with a smooth initial condition, obtained from running a simulation with well coupled dust and then perfectly coupled dust such that the initial vortices were smooth and axisymmetric, as this initial conditions have high initial dust the unstable phase is reached very rapidly in only a few shearing times. In all these cases the final outcome for the large vortices is the same. The vortices go unstable and start shedding small vortices, this process continues until what is left is a dust ring and many, smaller, stable vortices. This outcome can be seen in figure \ref{fig:DV-ring} where a large vortex (using $c_s=1.0$) has been allowed to evolve until it is completely broken up at $\Omega t=5000$.

\begin{figure}
    \begin{subfigure}{\columnwidth}
	    \includegraphics[width=8cm]{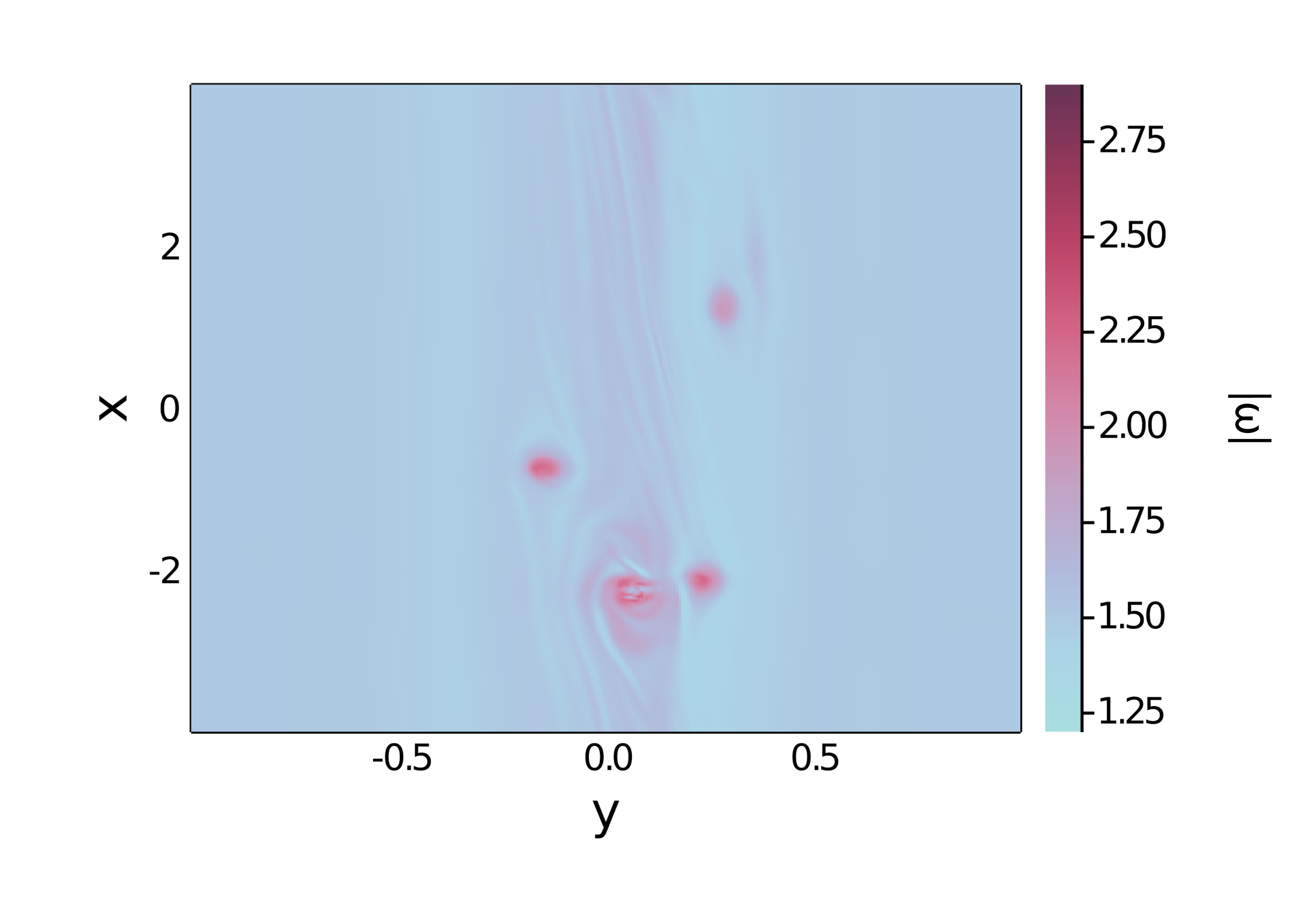}
	    \caption{Vorticity plot of the top right panel from figure \ref{fig:DV_Life-of-Vortex}, at $\Omega t = 4500$. The vortex has broken up. 
	    }
	    \label{fig:DV_v450_1}
    \end{subfigure}
    
    \begin{subfigure}{\columnwidth}
	    \includegraphics[width=8cm]{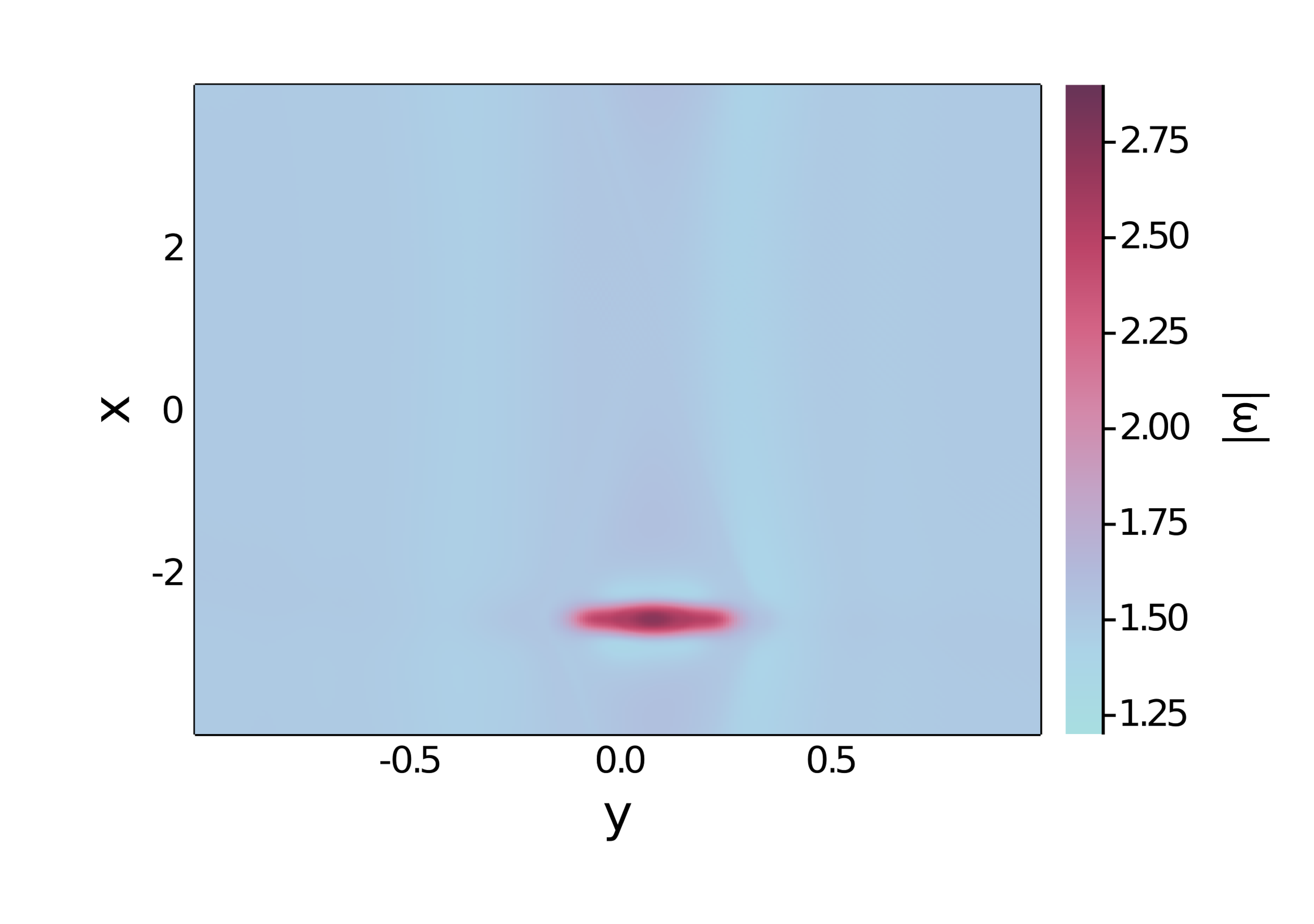}
	    \caption{Vorticity plot of the vortex in the bottom right panel from figure \ref{fig:DV_Life-of-Vortex}, at $\Omega t = 4500$. This
	    vortex shows no signs of instability. 
	    }
	    \label{fig:DV_v450}
    \end{subfigure}
    \caption{
    Vorticity plots of the final state of the vortices shown in figure \ref{fig:DV_Life-of-Vortex} at $t=4500\Omega^{-1}$. In panel \ref {fig:DV_v450_1} the dust feedback was not removed at $\Omega t=2250$ and the vortex proceeded to become unstable, breaking up into multiple smaller vortices.
    In panel \ref{fig:DV_v450} the dust cooling term was turned off at $\Omega t=2250$. After removing dust feedback at $\Omega t=2250$, demonstrating further that the instability is a drag instability mediated by the dust feedback and not the heavy core instability, which would not depend on feedback, only core dust concentration.
    }
\end{figure}
The breaking up of the vortex can also be seen in the vorticity evolution, see figure \ref{fig:DV_v450_1}. At its peak strength, the vorticity reached $\sim 2.75$, but at $\Omega t=4500$ the vortex has weakened considerably and broken up into several, smaller vortices. These smaller vortices fall into the category $b \ll H$, and if they are not disturbed by interaction with other vortices, they can survive for a very long time. 

The lifetime of the vortices strongly depends on the stopping time of the dust. Sufficiently well coupled dust ($\Omega\taus<10^{-4}$) only starts showing instability in large vortices after the dust to mass ratio in the vortex core starts approaching unity. For the highest stopping times, $\Omega\taus=10^{-2}$, a steadily contracting configuration was never reached and the vortices broke up immediately. No simulations at higher stokes numbers were run, as the terminal velocity approximation is known to perform poorly for $\Omega\taus>10^{-2}$ \citep{Lin_2017,Lovascio_2019}. 

\subsection{Turning off the dust feedback}
Especially interesting, is that the vortices can be brought back to a stable state after the instability has started occurring by setting the stopping time to zero, shutting off the cooling term. This is illustrated in figure \ref{fig:DV_Life-of-Vortex}, where, on the top row, the simulation is allowed to run its course and the vortex goes unstable and breaks up into smaller vortices, as described in the previous section. This can also be seen in the vorticity plot of the unstable vortex at $\Omega t=4500$ (figure \ref{fig:DV_v450_1}), showing the vorticity of the top right panel of figure \ref{fig:DV_Life-of-Vortex}. If instead the dust feedback on the gas is turned off when the instability first appears, the vortex readjusts to a stable configuration (figure \ref{fig:DV_Life-of-Vortex}, bottom row), remaining stable to the end of the simulation. The vorticity distribution of the stable vortex configuration at $\Omega t=4500$ is shown in figure \ref{fig:DV_v450_1}, which shows the vorticity of the bottom right panel of figure \ref{fig:DV_Life-of-Vortex}. What this means in practice is that the instability is not driven by the dust loading in the vortex, but by the dust drift.

\subsection{Vortex-vortex interaction} \label{sec:interactions}
When setting up the simulation, it is possible to chose a simulation domain that is too small. Due to the periodicity of the simulation domain, the vortex is able to interact with its counterparts across the periodic boundary, giving rise to different dynamics. This setup, while messy, is not unphysical: chains of vortices may form in protoplanetary discs due to hydrodynamic instabilities like the Rossby Wave Instability \citep{Lovelace_1999, Fu_2014}. When this vortex self interaction takes place a new instability arises, in vortices of all sizes. The interaction between vortices leads to the vorticity maximum to lead the pressure maximum. In this configuration the heavy core vortices are subject to the traditional Rayleigh-Taylor instability (RTI) as there is a pressure gradient in the opposite direction to the density distribution.  
\begin{figure}
\begin{subfigure}{\columnwidth}
	\centering
	\includegraphics[width=0.9\columnwidth]{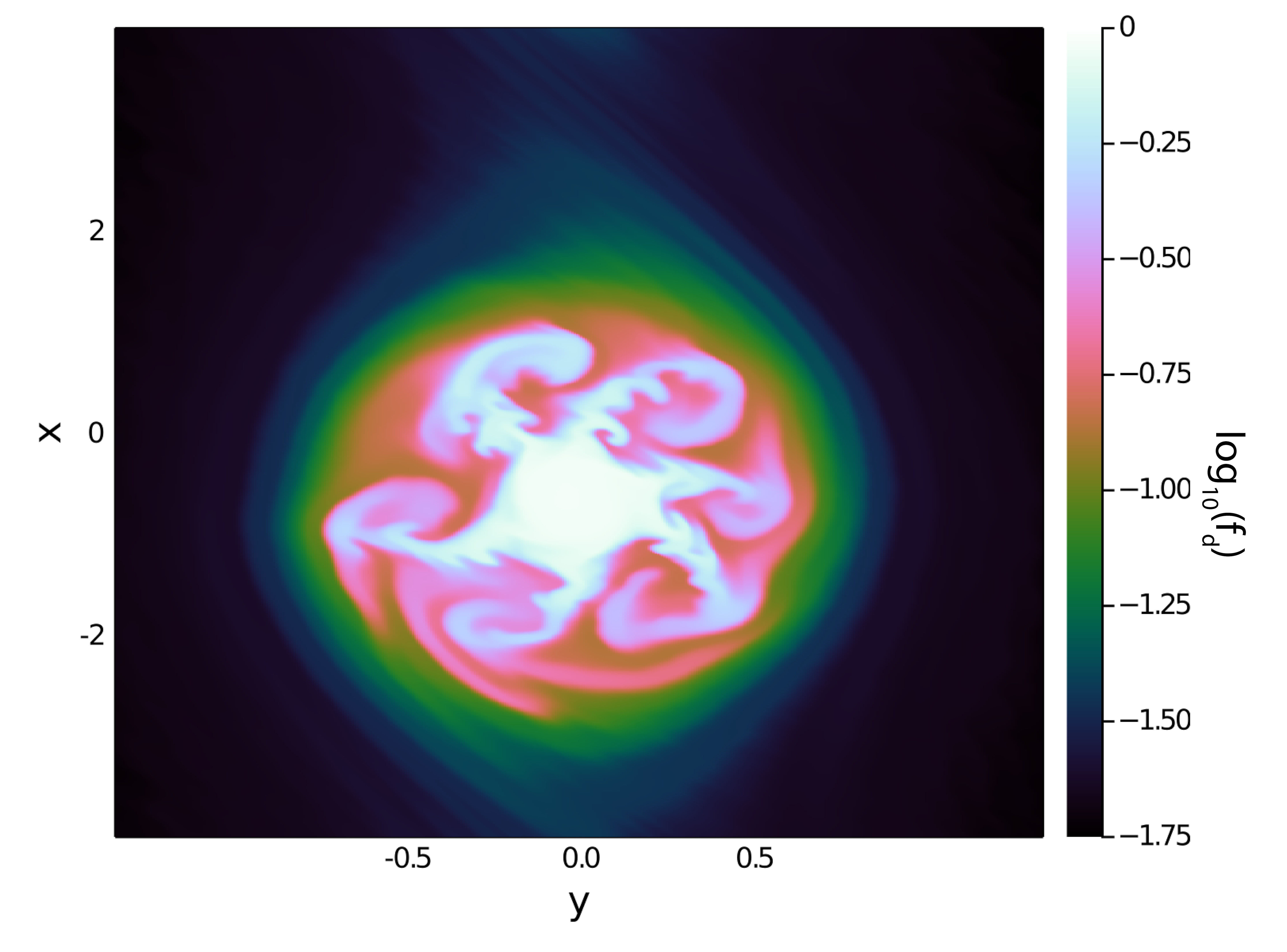}
	\caption{Dust to mass ratio of a vortex.}
	\label{fig:DV_fd95}
\end{subfigure}
\begin{subfigure}{\columnwidth}
	\centering
	\includegraphics[width=0.9\columnwidth]{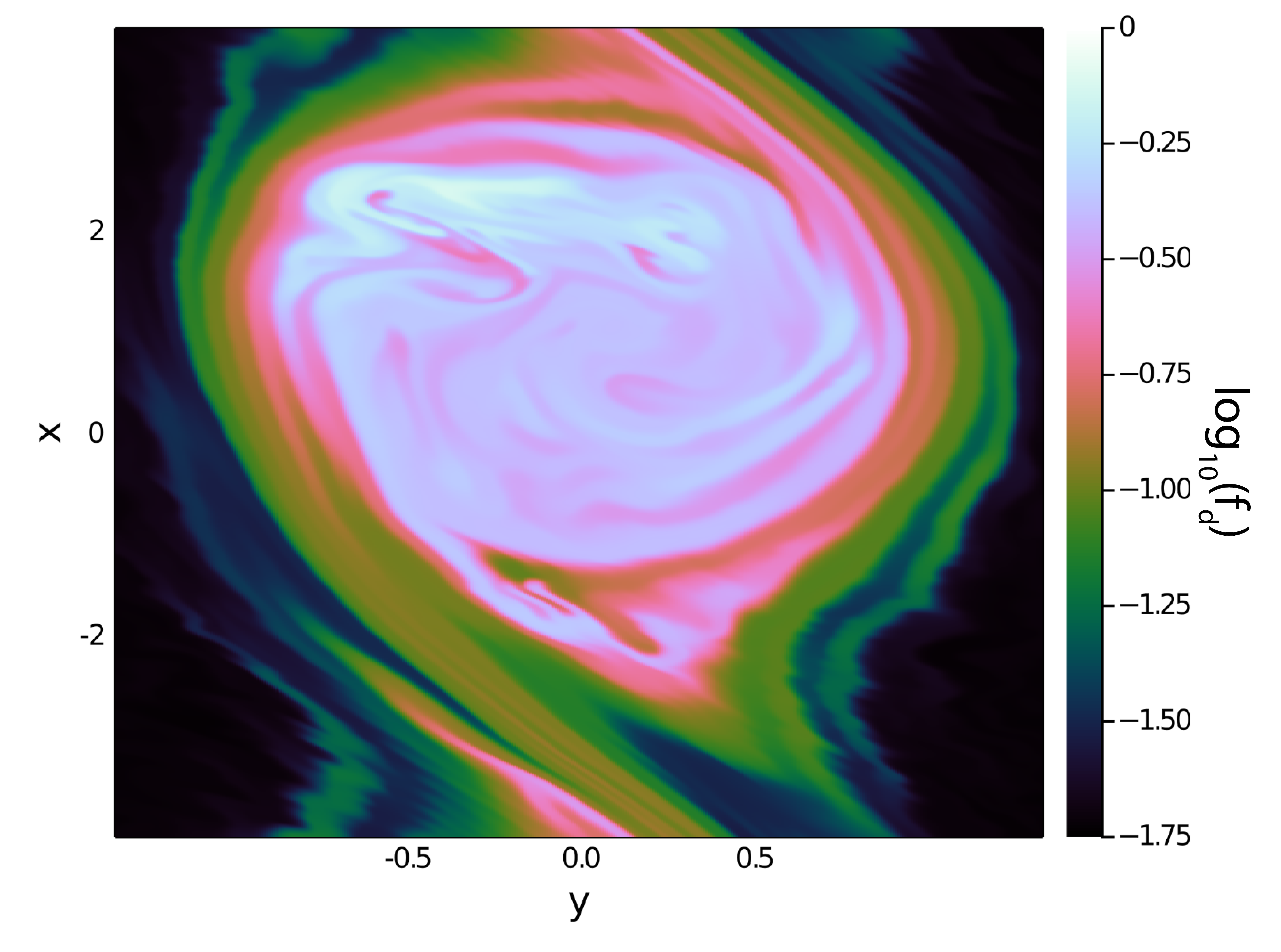}
	\caption{Dust to mass ratio of the vortex from panel \ref{fig:DV_fd95} 5 time-steps later.}
	\label{fig:DV_fd107}
\end{subfigure}
    \caption{When closely packed (azimuthal distance between vortices smaller than scale height), vortices go violently unstable. There is no longer a pressure maximum at the vortex core and the vortex rapidly breaks up due to a Rayleigh-Taylor like instability. The figure (panels \ref{fig:DV_fd95} and \ref{fig:DV_fd107}) shows the re-run with $\taus=0$ showing the instability is independent of dust drag. The instability mixes dust in the vortex rapidly, and eventually causes the vortex to break up into a dust ring. The figure shows the re-run with $\taus=0$ showing the instability is independent of dust drag.}
\end{figure}
This instability shows the hallmarks of the RTI, with high density tendrils rapidly extending into the low density region. The instability, at least in its early linear phase shows no dependence on dust coupling as the simulations were re-run starting from the pre-instability configuration with $\taus=0$ resulting in qualitatively the same outcome (dust drag acts to slightly smooth the density fronts of the high density tendrils, but otherwise the result was identical, the figures \ref{fig:DV_fd95} and \ref{fig:DV_fd107} both show the $\taus=0$ re-run). Runs with similar initial conditions, but no dust loading did not show this instability. These results are to be expected as the RTI rate only depends on the pressure and density gradients and cannot occur in the absence of a density gradient. This instability very rapidly destroys the vortex leaving only a dust ring in $<10$ shearing times. Some caveats should be kept in mind regarding these results, as this setup probably does not fully capture the physics of chains of tightly packed vortices in a disc for several reasons. The way that pressure is initially set up, in conjunction to the close packing, may contribute to creating the conditions for this instability to arise. It is also important to consider that by the way that the simulation is set up (in a periodic box), vortex mergers are prevented by effectively fixing the distance between vortices. This is especially important, as in global simulations vortices are observed to merge when in tightly packed chains. In practice this means that, while these simulations do not provide enough evidence to claim that in general chains of closely packed dusty vortices are unstable, they do certainly urge caution when setting up dusty vortices in periodic boxes.

\section{Discussion}
\label{sec:discussion}
Several results on dusty vortices have been obtained in previous works. These previous works consider the stability of large vortices, working both in shearing boxes \citep{Crnkovic-Rubsamen_2015} and in global simulations \citep{Surville_2019, Fu_2014}. Large dusty vortices are found to be unstable when loaded with dust. Both \citet{Surville_2019} and \citet{Crnkovic-Rubsamen_2015} find the instability to be dependent on the dust grain size. This result is in agreement with the results shown in this paper for large ($b \approx H$) vortices. Two hypotheses are invoked to explain the origin of the dusty vortex instability; the heavy core instability \citep{Chang_2010}, as suggested by \citet{Fu_2014} and \cite{Crnkovic-Rubsamen_2015}, and an unspecified drag instability, as suggested by \citet{Surville_2019}. Our results agree with the latter. In our simulations, large dust-laden vortices go unstable late in their lifetimes, after having collected large amounts of dust, enough to make their core very dust dominated. However, these vortices can be brought back from instability by changing the dust stopping time to zero, showing that the instability is driven by drag rather than the dust loading itself. On the other hand, small dusty vortices did not go unstable within the time-frame of our simulations. 

\subsection{Understanding the source of the instability}
In this paper we establish that the radial size of dusty vortices is an important factor for their stability. Our simulations showed no evidence for the existence of the heavy core instability. The heavy core instability arises in vortices, when sufficient density contrast between the vortex core and envelope has formed \citep{Sipp_2005}. This instability is a hydrodynamic instability akin to the Rayleigh-Taylor instability in a rotating frame of reference \citep{Chang_2010}. The driver behind the instability is the density gradient of the fluid in the opposite direction to the centrifugal force. It can therefore be concluded that, the heavy core instability should not depend on dust coupling. In fact, it should operate even for perfectly coupled dust. This property of the heavy core instability allows for us to test whether the vortex instability observed is indeed the heavy core instability by disabling the dust coupling term in an unstable vortex configuration. This amounts to simulating perfectly coupled dust, which should still be unstable to the heavy core instability. Knowing that the setup is unstable when considering dust-gas coupling, means that if the instability does not appear in the perfectly coupled ($\taus=0$) case the heavy core instability can be ruled out as the instability observed in the $\taus>0$ case. In our simulations, turning off the cooling term in a vortex once unstable killed off the instability (figure \ref{fig:DV_Life-of-Vortex}). This means that the dust cooling term, and thus dust-gas drag, drives the instability, rather than just the centrifugal force acting against the density gradient. This rules out the heavy core instability and suggests a drag instability. The instability observed in our simulations seems to be the same one observed by \citet{Surville_2019}, and our results support their conclusion that it is a drag instability breaking up the vortices. More work is required to understand why the instability appears to vanish in smaller vortices, though, the dependence on vortex size or sound speed can offer a hint regarding the exact nature of the instability. 

In our simulations we do not observe the heavy core instability and have ruled it out as the source of instability for our large vortices. Our results do not, however, rule out the heavy core instability in general because, as discussed by \citet{Railton_2014}, the issue with the heavy core instability is, that for it to be able to act, the fluid must be shear-free. Vortices in our simulations always possess a significant amount of shear so that the heavy core instability can not operate. In general, vortices in astrophysical settings are not shear-free, with notable exception vortices in Keplerian discs with aspect ratio $q=7$ \citep{Railton_2014}. The way we setup our simulations makes it impossible to create a shear-free vortex. While this would be possible using the methods described in \cite{Railton_2014}, when the dust is not perfectly coupled the vortex will move away from this initial condition due to gas drag. As the dust drifts towards the vortex core, it changes the shape of the vortex, this means that setting up a shear-free vortex is not straight forward and it is not clear that the shear-free configuration would be long lived enough for the heavy core instability to grow to the point where it destroys the vortex.  

\subsection{Vortex lifetime and implications for observations and planet formation}
Increasingly high resolution imaging of protoplanetary discs has only recently been able to observe the first small scale non axisymetric features in protoplanetary discs \citep{DSHARP_2018,Booth_2021,van-der-Marel_2021,Tsukagoshi_2019}. Given that vortices have plenty of opportunities to form in a disc, these vortices must either be too short lived, too small to be observed with current telescopes, or be very uncommon for other reasons. Currently only one instance of small scale non-axisymmetric structure has been observed \citep{Tsukagoshi_2019}, but the origin of the over-density is unclear, with leading theories being a forming planetesimal and a dust rich vortex core. The long lifetime of small vortices in this study poses no issue with current observations, as the small, stable vortices with radial extents much smaller than the scale height of the disc would not be visible in observations. 
The larger unstable vortices with radial extents of order the scale height of the disc pose more questions. The lifetime of these vortices is limited by the drag instability which appears as dust is concentrated in the vortex core, but the lifetime of the vortex is nonetheless of order many hundred shearing times, especially for smaller dust grains. This means that the larger vortices are able to concentrate large amounts of dust at their cores and survive for many orbits before breaking up. It may be feasible to observe vortices of size similar to the scale height with current or upcoming technology. To verify whether these dusty cores are observable radiative-transfer simulations with synthetic observations are required.

The dust collecting capacity of vortices in protoplanetary discs has exciting implications. Small to medium sized ($b\ll H$ to $b\sim H$) vortices in discs survive at least long enough to bring about large increases in local dust to gas ratio. This increase in local dust to gas ratio can drive faster dust growth, due to coagulation through increased collision frequency. Even the medium sized vortices, which break up after a few hundred orbits may aid planet formation after their breakup, by leaving behind a dust enhanced ring making the SI more active in the region. Increasing background dust to gas ratio is known to speed up the SI \citep[e.g.]{Li_2021}, but also provides more material to clump and making gravitational collapse of the clumps more likely. The change in distribution of dust sizes due to variable dust drift rate with dust particle size in the vortex may also be important for driving SI. In \citet{PSI3} it was shown that altering the dust distribution, for example increasing the concentration of large grains compared to the smaller grains, can have a huge impact on the growth rate of polydisperse drag instabilities. This of course raises the question, how does polydisperse dust affect the evolution of vortices? Answering this question in detail requires a nonlinear polydisperse or at least multiple dust species solver and a large amount of computer time. Solving the polydisperse dust equations non-linearly, in fact acts to add an extra dimension in dust grain size space, making 2D simulations approximately as computationally expensive as 3D ones and 3D simulations even more so. Such a study of polydisperse dust in vortices could produce interesting results, This is because, recent work \citep{PSI1,PSI3} has shown polydisperse dust to damp drag instabilities, suggesting that vortices may be more stable in a polydisperse dust than a monodisperse dust. Overall adding polydisperse dust to vortex simulations would make the simulations still more realistic and allow more detailed studies of the effects of vortices on dust mass and size distribution in discs.

\subsection{Forming vortices self consistently}
Several vortex forming mechanisms can occur in PPDs. To self consistently form vortices in simulation, the conditions for one of these mechanisms to occur need to be set up. This approach has both advantages and drawbacks. For this study, using self consistent vortex formation would have limited the scope of the study, as it limits the control on vortex parameters, for example the RWI can only form large vortices with radial extents similar to the scale height. Different avenues of vortex formation can lead to the formation of vortices with  different parameters as well as different numbers of vortices. The results in this paper, suggest that many configurations are likely unstable, notably tightly packed vortex trains. In tight chains, the vortices end up with a pressure minimum at their cores, leading to strong Rayleigh-Taylor type instability. The mechanism that leads to this is unclear, but it points towards the likely instability of tightly packed vortex chains. Vortex formation mechanisms that form fewer, more spaced out vortices, are more likely to form vortices that will survive long enough to collect large amounts of dust in their cores. This, however, does not mean that all vortices in a chain must go unstable. To test whether this is the case, however, global simulations of dusty discs are required. This adds a level of complexity to the study of vortices in dusty protoplanetary discs, but is important for understanding the role vortices have to play in protoplanetary discs.

\subsection{Vortices in 3D}
Protoplanetary discs are three dimensional: they have a finite height and considering the height dimension can change the dynamics observed in a simulation of protoplanetary discs. Disc winds acting higher in the disc for example can affect the evolution of the disc as a whole by driving momentum transfer through the disc \citep{Suzuki_2016}. This applies to vortices in discs too. Vortices have been shown to have much shorter lifetimes in 3D than in 2D \citep{Lesur_2016} this is because in 3D vortices become subject to elliptic instabilities which destroy the vortex \citep{Railton_2014}, pebble back-reaction though may not directly affect this \citep{Lyra_2018}. Small vortices are known to be subject to such instabilities, which puts into question the true lifetime of small dust loaded vortices in PPDs. Overall, several questions remain to be answered in regards to the lifetime and stability of dust loaded vortices in 3D PPDs. There are also questions to be answered with regards to the vertical dust distribution in 3D vortices, like what are the effect of a stratified disc on the evolution of a small to medium sized vortex. Follow-up work to this paper will analyse the role dust plays in the evolution of vortices in 3D, attempting to answer whether the dust collecting potential of vortices is maintained in 3D and whether dust acts to stabilise or further destabilise vortices in 3D.

\section{Conclusions}
\label{sec:conclusion}
We have presented two-dimensional, local simulations of vortices in protoplanetary discs containing both gas and dust. Due to the pressure maximum at their centre, dust collects inside the vortices, leading to dust-dominated vortex cores. For vortices with semi-minor axis much smaller than the scale height of the disc, this configuration was found to be stable over at least $\sim 1000$ local orbital periods. Bigger vortices, with semi-minor axis similar to the scale height of the disc, were found to be prone to a drag instability that destroys the vortex and leaves an axisymmetric ring of dust. The sizes and aspect ratios of vortices in protoplanetary discs depend on the formation mechanism. Planet-disc interactions tend to form large vortices with semi-minor axes larger than the local disc scale height. Other mechanisms like the VSI can form smaller vortices of size similar to the scale height or smaller, like the ones discussed in this paper. Additionally, the VSI does not require the disc to already host planets, making it perhaps more relevant to planetesimal formation. The results shown in this paper suggest that small to medium size vortices, even if ultimately unstable, can cause sufficient dust enrichment to aid planet formation, small and medium sized vortices may therefore play a key role in planet formation. Future work should consider whether these results hold in three dimensions.   

\section*{Acknowledgements}
We acknowledge that the results of this research have been achieved using the DECI resource Beskow based in Sweden at PDC with support from the PRACE aisbl.
This research utilised Queen Mary's Apocrita HPC facility, supported by QMUL Research-IT \citep{apocrita}.
FL was supported by an STFC studentship.
SJP is supported by a Royal Society URF.

\section*{Data Availability}
All simulation setups will be made available on request to the authors. Data to reproduce the figures can also be provided. 



\bibliographystyle{mnras}
\bibliography{Bib} 


\appendix


\bsp	
\label{lastpage}
\end{document}